\documentclass []{aa}
\usepackage{psfig}
\begin{document}
\thesaurus{11.06.2; 11.07.1; 11.09.2; 11.11.1; 11.19.2}
\title{The Intra-cluster Medium Influence on Spiral Galaxies}

\author{C.~Adami \inst{1,2}, M.~Marcelin \inst{3},
P.~Amram \inst{3}, D. ~Russeil \inst{3,4}}

\institute{
IGRAP, Laboratoire d'Astronomie Spatiale, 
Traverse du Siphon, F-13012 Marseille, France  
\and Dearborn Observatory, 
2131 Sheridan, 60208-2900 Evanston, USA 
\and IGRAP, Observatoire de Marseille,
2 Place Leverrier, 13004 Marseille France
\and AAO, Coonabarabran, NSW 2357, Australia
}
\offprints{C.~Adami} 
\date{Received date; accepted date} 

\maketitle 
 
\markboth{The Intra-cluster Medium Influence on Spiral Galaxies}{} 
 
\begin{abstract} 

We made a detailed analysis of the sample of 39 cluster spiral galaxies of 
various 
types observed at H$\alpha$ wavelength by Amram et al. (1992 to 1996), 
with a scanning Fabry-Perot. We
plotted the outer gradient of their rotation curves as a function of the 
deprojected cluster-centric distance. 
The rotation curves of late type galaxies markedly rise far from the cluster 
center. This suggests evolutionary effects, 
since early types show no special trend. 
We suggest that the evolution process within a cluster 
leads late type galaxies to exhibit flatter curves when they get closer to 
the center, on their way to evolving into early type galaxies.

\end{abstract} 
 

\section{Introduction}

The study of cluster galaxies helps us understand and constrain models
of galaxy-evolution.
Recently, several authors have used samples of thousands of galaxies to 
outline the main global properties of galaxies in clusters (see Adami et al. 
1998 and references therein) such as the relation between the galaxy type with 
its environment or with its cluster centric distance.
Some other studies (Rubin et al. 1988 and Whitmore et al. 1988: RWF hereafter,
Amram et al. 1992 to 1996: hereafter AmI, II, III, IV and V or Sperandio
et al. 1995: Sp hereafter) have made a more precise analysis (although with 
smaller samples) of the spiral morphological types to determine
if their rotation curves are related to the cluster centric 
distance. 
However, these last studies suffer from an evident bias: they are based on a 
bidimensional analysis. Projection effects could explain the 
contradictory results among the authors.
We analysed the sample of 45 galaxies observed by 
Amram et al. (AmI, AmIII and AmIV) in various clusters and used a
statistical deprojection technique (based on the density profile of each of 
the spiral morphological types) to extract the main properties of the
spiral galaxies in clusters and to constrain the models.
In section 2 we describe the samples and the methods to discard
interloppers. Section 3 describes the deprojection technique and 
the results. We discuss the implications in the last section.

\begin{table*} 
\caption[]{Spiral galaxies of the total sample. Col.(1) gives the name
of the cluster, col.(2) the name of the galaxy, col.(3) the value of OG, 
col.(4) the projected cluster centric distance: d2D, col.(5) the likely 3D 
clustercentric distance when possible: d3D, col.(6) the statistical 1 $\sigma$ 
error on d3D, col.(7) the likely 3D clustercentric distance when possible
computed with only one random selection: d3DII, col.(8) the RC3 morphological
type and col.(9) the origin of the data (AmI, III or IV).}
\begin{flushleft} 
\begin{tabular}{ccccccccc} 
\hline 
\noalign{\smallskip} 
Cluster & Gal. name & OG ($\%$) & d2D(Mpc) & d3D(Mpc) & $\sigma$d3D(Mpc) &
d3DII(Mpc) & Morph. type & Authors ref.\\ 
\hline 
\noalign{\smallskip} 
A0262 & NGC668* & -2 & 2.14 & 2.97 & 0.66 & 3.03 & 3 & AmIII \\
 & NGC669 & 7 & 2.17 & 2.58 & 0.70 & 3.78 & 2 & Am III \\
 & NGC688* & 0 & 1.66 & 2.66 & 0.84 & 2.82 & 3 & AmIII \\
 & NGC753* & -5 & 1.33 & 2.43 & 0.75 & 2.39 & 4 & AmIII \\
 & UGC1347* & 5 & 0.32 & 1.52 & 0.97 & 1.59 & 5 & AmIII \\
 & UGC1493* & -1 & 2.88 & 3.41 & 0.47 & 3.25 & 2 & AmIII \\
A0539 & UGC3269* & 7 & 1.20 & 2.33 & 0.80 & 3.51 & 4 & AmI \\
 & UGC3282* & 5 & 0.86 & 2.06 & 0.89 & 2.48 & 6 & AmI \\
A1367 & NGC3861 & -9 & 0.37 & 1.64 & 0.99 & 0.45 & - & AmIII \\
 & NGC3883* & -2 & 1.65 & 2.60 & 0.66 & 2.45 & 3 & AmIII \\
 A1656 & NGC4921 & 0 & 0.76 & 1.97 & 0.92 & 0.81 & 2 & AmIII \\
 & UGC8161* & 12 & 2.72 & 3.32 & 0.40 & 2.93 & - & AmIII \\
 & Z130-008* & 21 & 4.90 &  &  &  & 6 & AmI \\
 & Z160-058* & 19 & 2.90 & 3.42 & 0.31 & 3.34 & 5 & AmI \\
 & Z160-106 & 35 & 1.15 & 2.29 & 0.81 & 3.72 & -2 & AmI \\ 
A2151 & IC1179 & 5 & 0.14 & 1.38 & 1.05 & 1.12 & - & AmIII \\
 & NGC 6045 & 10 & 0.14 & 1.33 & 1.04 & 0.43 & 5 & AmI \\
 & NGC 6050 & 1 & 0.14 & 1.35 & 1.04 & 0.49 & 5 & AmIII \\
 & NGC6054* & 3 & 0.17 & 1.36 & 1.04 & 0.56 & 3 & AmIII \\
 & UGC10085* & 8 & 7.30 &  &  &  & 6 & AmI \\
DC1842-63 & DC 10* & 32 & 0.11 & 1.30 & 1.05 & 0.63 & - & AmIV \\ 
 & DC24* & 8 & 0.15 & 1.34 & 1.04 & 0.72 & 4 & AmIV \\
 & DC39* & -3 & 0.19 & 1.38 & 1.04 & 1.31 & 3 & AmIV \\
 & DC47 & 13 & 0.31 & 1.51 & 1.02 & 1.47 & - & AmIV \\
 & WR42 & 5 & 0.89 & 2.08 & 0.88 & 1.66 & 2 & AmIV \\
Cancer & NGC2558* & 2 & 0.69 & 1.90 & 0.93 & 1.86 & 2 & AmI \\
 & NGC2595* & 6 & 1.90 & 2.82 & 0.60 & 2.10 & 5 & AmI \\
 & UGC4329* & -5 & 0.44 & 1.67 & 0.99 & 1.09 & 6 & AmI \\
 & UGC 4386 & 24 & 0.91 & 2.09 & 0.88 & 1.54 & 3 & AmI \\
 & Z119-043* & 2 & 0.47 & 1.68 & 0.98 & 0.95 & - & AmI \\
 & Z119-051* & 22 & 0.43 & 1.64 & 0.99 & 1.05 & - & AmIII \\
 & Z119-053* & 6 & 0.32 & 1.52 & 1.02 & 1.16 & - & AmI \\
Pegasus I & NGC7536* & 13 & 5.20 &  &  &  & 4 & AmI \\
 & NGC7591* & -7 & 1.62 & 2.63 & 0.68 & 2.13 & 4 & AmIII \\
 & NGC7593* & 8 & 3.00 & 3.48 & 0.29 & 3.40 & 5 & AmI \\
 & NGC7631* & 4 & 0.29 & 1.49 & 1.02 & 0.43 & 3 & AmI \\
 & NGC7643* & 5 & 3.64 & 3.82 & 0.38 & 3.89 & 5 & AmI \\
 & UGC12498* & -7 & 0.26 & 1.45 & 1.03 & 0.56 & 3 & AmI \\
 & J2318+0633* & 5 & 1.62 & 2.63 & 0.68 & 3.84 & - & AmIII \\
\noalign{\smallskip} 
\hline	    
\normalsize 
\end{tabular} 
\end{flushleft} 
\label{} 
\end{table*}

We have assumed here H$_0$=75 km.s$^{-1}$Mpc$^{-1}$.

\section{The sample}

We use here the shape of the Rotation Curve (RC hereafter) of spiral galaxies 
to study
the effect of the intra cluster medium on the halos of these galaxies.
Amram et al. (AmI, AmIII and AmIV) observed 45 cluster galaxies with a scanning 
Fabry-Perot interferometer and with 3.6m telescopes
(CFHT or ESO) at H$\alpha$ wavelength. They drew detailed velocity fields
with high resolution (both spectral and spatial) and derived the RC with
a much better precision than can be done with slit spectroscopy. A detailed
comparison of both techniques is given in AmI.

The shape of the outer part of the RCs is given by the outer gradient "OG", 
defined as the difference
between the velocity at 0.4R$_{25}$ and 0.8R$_{25}$, normalized to the maximum
velocity of the RC. R$_{25}$ is the optical radius of a galaxy defined to
the point where the surface brightness is 25 mag/arcsec$^2$. 

From the total sample of 45 galaxies Amram et al. could measure 39 OG (AmV).
In some cases, the OG were obtained by a slight extrapolation of the RC.
These 39 galaxies are listed in Table1, with the corresponding OG and the distance parameters that are discussed in section 3.

Although the OG is perhaps not the best parameter to define the tendency of
a rotation curve to rise or decrease beyond the optical radius it has been
used by many authors and appears as a common language. Also it is easy to
compute as soon as the RC of a galaxy is available.

In order to get a sample as clean as possible we decided to discard the  
galaxies for which the OG was obtained with too large an extrapolation. We only 
kept those for which the RC reached at least 0.7R$_{25}$. 
We thus excluded WR 42 and NGC 4921. The original data for NGC4921 had been
obtained with rather bad weather 
conditions and were of rather poor quality, providing a RC with large error 
bars, and the OG was found abnormaly large (22). Although, we re-observed it
at CFHT in 1995 in better conditions, with the same instrument, and find now a 
flat curve with OG close to zero. However it still remains the result of an 
extrapolation since the RC barely reaches 0.6R$_{25}$.

We also discarded galaxies with a high extinction: we
thus removed the galaxies with an inclination greater than 75 degrees, namely
NGC 669 and UGC 4386 (both having an inclination of 80 degrees).

We also removed interacting galaxies to avoid effects due to the interaction
on the RC. Six galaxies were thus excluded (see appendix for details).

\begin{figure*} 
\vbox 
{\psfig{file=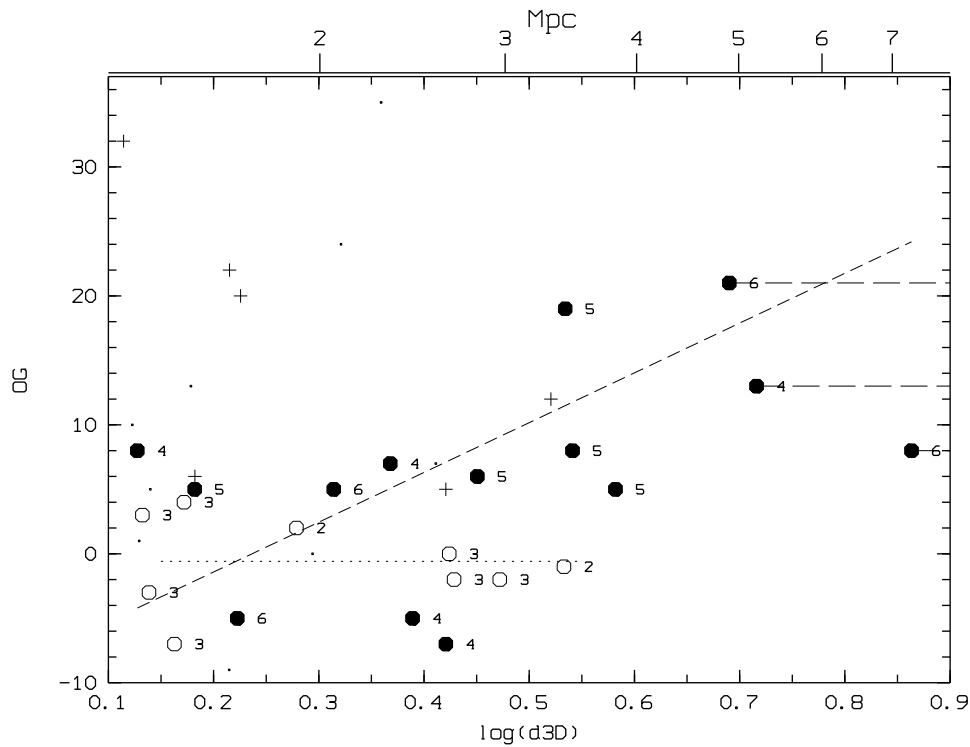,width=15.0cm,angle=270}} 
\caption[]{Distribution of the OG values (in percents) with the decimal 
logarithm of the 3D cluster-centric distance (in Mpc). The 39 galaxies from
AmV are plotted, with Points for those ignored in the discussion because of 
selection criteria (mainly because of inclination or interaction). 
Open circles are for 
early type galaxies, with corresponding types 2 or 3 mentioned aside. Filled
circles are for late type galaxies, with 
values 4, 5 or 6. Crosses are for undefined types. Regression lines are drawn
for both classes of galaxies, with a dotted line for early types and a 
dashed line for the late types ; the distances for the 3 galaxies at right, 
beyond 4Mpc, are lower limits as suggested by the dashed horizontal lines} 
\label{} 
\end{figure*}

This severe selection finally limits the total sample to 29 galaxies marked
with a * in Table 1. They were homogeneously selected in 8 nearby distinct 
clusters and have morphological types lying between 2 and 6 (following the
RC3 classification : de Vaucouleurs et al. 1989) although six of them have 
no clearly defined type. 

In this respect the case of DC 10 deserves an explanation. Although
it is refered to as a type 1 galaxy in the RC3 its type is most probably
around 4, as suggested in AmIV, and we decided to attribute no definite type
to this galaxy. Furthermore its OG is abnormally high (32), although it 
successfully passed 
all the selection criteria, placing it clearly outside our average distribution
of points on Fig.1 (its d2D and d3D values are respectively 0.11 and 1.3 Mpc).
We have no explanation to this, but we suppose that this galaxy
suffers from projection effects that are impossible to correct due to its 
apparent position very close to the center of its cluster.

We also checked the projected clustercentric distances
given by the authors by using an X-Ray determination of the center whenever 
possible (for A262, A539, 
A1656, A2151 and Pegasus). The difference between the centers used by the
authors and the X-ray centers are quite small (typically less than 0.1 Mpc)
and may be neglected regarding the cluster centric distances used here
(see Tab.~1).

\section{Other samples of Rotation Curves}

We looked through the literature for other RCs of cluster galaxies, in order
to extend our number of OG values. The other main samples of RC are all obtained
through slit spectroscopy. We now discuss these samples.

Mathewson et al. (1992) :  
Of 965 spiral galaxies for which they give H$\alpha$ RC only 261 belong to 
clusters. 
Applying our selection criteria (i.e. removing too much inclined galaxies, 
interacting ones and those with RC within 0.7R$_{25}$) we keep fewer than 
100 galaxies. Many of the remaining galaxies exhibit RCs with strong asymetries 
or dispersions and are not suited for obtaining reliable OG values. 
Furthermore most of them are found in loose clusters, not useful for the study 
of environmental effects. At last we deal with fewer 
than 10 galaxies in rich clusters (some of them being already found in our own 
sample: AmI + AmIII + Am IV) and we decided not to use this sample.

Mathewson and Ford (1996) :
They added 1051 H$\alpha$ RC of galaxies to the previous sample. Applying the 
same selection criteria, we keep fewer than 40 galaxies. Furthermore it was
hard to get reliable values of OG from the published curves and we decided 
not to use this sample for our paper.

Persic and Salucci (1995) :
This is a subsample of Mathewson et al. with 80 high quality RCs of galaxies.
Most of them are field galaxies and, after rejection of that are too inclined,
it remains only a few galaxies in loose clusters. Hence, we did not use this
sample.

Courteau (1997) :
The sample contains 304 field galaxies from the UGC catalogue, but none in 
clusters. 

Corradi and Cappacioli (1991) :
This catalogue of kinematical data about 245 galaxies is based on kinematical
studies found in the literature. It is not a compilation of RCs but provides an
interesting classification of the shape of the RC around the optical radius,
with three classes : rising, flat and decreasing. We use these data in
section 5 (discussion and conclusion).

Dale et al. (1997,1998) :
Dale published a huge sample of 522 RC of late-type cluster galaxies in his 
PH D thesis. The data, now completed by the morphological type information, 
are being prepared for publication. The two first papers that are already 
published contain
145 RCs of cluster galaxies, of which 89 remain after removing higly inclined 
ones and those with abnormal radial velocities.
The use of these RC is not straightforward however, since no R$_{25}$ is 
available for these galaxies (the parameters given by the authors are 
R$_{23.5}$ and R$_{83}$, radius of the isophote containing 83\% of the 
total emission in the I band). So that there is no direct way for getting OG
from these data. Moreover, some of the last 89 galaxies have a RC not enough
extended to allow a determination of OG.

Almost all of the samples listed above were intended to check the Tully-Fisher 
relation.
Let us now examine the samples specificaly devoted to the study of OG in 
cluster \mbox{galaxies :}

Rubin et al. (1988) and Whitmore et al. (1988) :
They measured 16 OG, 10 for early type galaxies (Sa + Sb) and 6 for late type
galaxies (Sc). Most of them were measured with more accuracy, owing to the
scanning Fabry-Perot, by AmI and AmIII. Only 2 galaxies satisfying our 
selection criteria could be added to the sample discussed in this paper,
namely, UGC 12417 and WR 66.
 
Distefano et al. (1990) :
They measured 15 OG of galaxies, most of them in the Virgo cluster. For the 9 
OG values that were obtained by a combination of H$\alpha$ and HI data, the
optical RC could not be drawn far enough and we discarded them. 
Finally there are only 5 galaxies from this sample (1 early and 4 late)
satisfying our selection criteria : NGC 4254, 
NGC 4294, NGC 4501, NGC 4651 and NGC 4654.

Sperandio et al. (1995) :
They measured 16 OG of galaxies in the Virgo cluster (2 early and 16 late),
4 of which (2 early and 2 late) satisfy our selection criteria : NGC 4178, 
NGC 4480, NGC 4639 and NGC 4713.

From the 3 above samples, according to our selection criteria, it is 
possible to add 11 galaxies to our own sample 
of 29 galaxies from Fabry-Perot observations discussed in section 2.
We did not plot them on Fig.~1 for sake of homogeneity (the 11 extra galaxies
having been observed through slit spectroscopy) but it is worth noting 
already that these addition data
reinforce the conclusions reached from the analysis discussed 
hereafter (see end of section 4).

\section{Analysis}

\subsection{Deprojection}

We have used the results of Adami et al. (1998) to find a way to deproject
statistically the cluster-centric distances of our sample of galaxies. The
method is based on the distribution profile of the spiral galaxies. Using a 
very large sample of about 2000 galaxies in 40 clusters, they have shown that
the Sa+Sb galaxies (resp. Sc+Sd+Sm+Irr) follow a King profile with a core 
radius of 0.212 Mpc (respect. 0.263 Mpc). We have assumed these profiles for 
the present galaxies and via the Abel inversion, we have statistically 
deprojected these King profiles (they have the same core radius in 2 or 3 
dimensions). We note that the values we get are only statistical estimations 
and must not be interpreted as reliable values for individual galaxies.

For each galaxy we have made 10$^6$ random generations of the spatial 
cluster-centric distance according to the given 3D King density law. We have
kept only the distances greater than the observed projected distance 
(the projected distance observed in the sky plane being always lower than 
the 3D distance). The dispersion of these distances gives an estimation
of the error we have for each galaxy (according to the size of the error,
some of our galaxies are consistent with a central location in the
cluster). 

Finally, we have imposed a maximum radius for the cluster equal to 4 Mpc. This 
is about the maximum virial radius observed for the clusters (see e.g. Carlberg
et al. 1996). A larger distance would imply that the isothermal sphere model 
(and then the deprojection method) is not valid.

We note that we computed here a value similar to the analytical mean 
weightened with a
truncated King profile (due to the large number of realizations). Two
galaxies with the same projected cluster-centric distance will have
the same deprojected distance. To take into account this degeneracy, we
have used a second deprojected distance: a single value generated by a 
random generator and weighted for the King profile (taking into account the 2D
constraint), as suggested by the referee. This gives different values of the 
deprojected distance for a given projected distance. This allows to evaluate
the robusteness of the 2 methods.

In Table 1 we give, for each galaxy, the mean likely 3D distance from the 
center of the cluster (hereafter d3D) computed from the observed projected 
distance in the sky (d2D), the second estimation of the 3D distance (d3DII) and
the error we have for d3D ($\sigma$d3D).

The adopted maximum radius of 4 Mpc led us to not apply any deprojection to 
the galaxies with projected distances greater than this radius : 
they were considered to be field galaxies in the present paper (this 
is the case for the 3 outermost galaxies of our sample).

\subsection{Results}

Fig.~1 shows the variation of OG as a function of d3D for the 39 galaxies
of AmV having a measured OG. The plot is the same as in Fig. 13 of AmV with 
d3D instead of d2D (only one OG value has changed in the meantime, that for
NGC 4921 which is now at 0 from the new data obtained at CFHT). 

AmV found no correlation between OG and the distance to the cluster center. 
Using the d3D distance does not change this result
since, as explained below, it may only amplify an already existing gradient.

With the selected sample of 29 galaxies following the selection criteria 
discussed in section 2 (that is to say discarding the galaxies plotted
as simple points on Fig.1) there is still no clear tendency of OG to vary
with the distance, especially since the dispersion is quite large.

Discarding now the galaxies with no defined type (23 galaxies remaining, 
with open and filled circles) things become clearer and there is a marked
tendency of OG to increase with the distance.

OG is found to increase as a function of distance following the relation:

	OG = (35.8$\pm$20.8)$\times$log10(d3D) + (-8.9$\pm$18.3)

N.B. This linear relation uses the bissector method from Isobe et al. 1990. The
value of the Spearman's rank correlation (Press et al 1992) is 0.11.

Using the d3DII estimator, we get consistent results with however a lower
value of the slope: (23.2$\pm$21.9). The two results being consistent, we have
choosen to use hereafter only the d3D statistical estimator.

Using the more usual projected distance d2D, instead of the corrected 
spatial distance d3D, one finds :

	OG = (12.6$\pm$9.8)$\times$log10(d2D) + (3.1$\pm$8.7)

The measured slope is much steeper when using the deprojected distance (about
3 times). This is because the deprojection method significantly increases 
the distance of galaxies close to the cluster
center, thus shortening the distance range and amplifying any gradient
of OG as a function of distance. We remark that the deprojection 
method, when applied to galaxies at the same projected distance, will place 
late types further than early types, although the difference is not significant
(for instance this difference is less than 3 \% for d2D = 2 Mpc, whereas it 
would be more than 10\% between ellipticals and late types).

The slope found here with the 2D distances (with the sample limited
to galaxies of well defined type) is significant,
although it is much smaller than the slope originally
claimed by RWF (our slope is about three times smaller than their value).

Let us now look at the behaviour of galaxies depending on their type :

On Fig.~1 one can see that OG remains constant for early type galaxies, 
whatever their location within the cluster, meanwhile there is a marked 
tendency of OG to increase with the distance to the cluster center with
late type ones. 

The linear relation found for early type galaxies (bissector method from 
Isobe et al. 1990) is (with the Spearman's rank correlations respectively 
equal to 0.21 and 0.19, indicating a better correlation):

OG = (-0.10$\pm$0.14)$\times$log10(d3D) + (0.55$\pm$3.3)

and that for late type ones:

OG = (38.6$\pm$18.4)$\times$log10(d3D) + (-9.13$\pm$16.2)

Both relations are plotted respectively as a dotted line and a dashed line 
on Fig.~1.

N.B. The 11 galaxies (4 early and 7 late) discussed at the end of 
section 3 (observed through slit spectroscopy by other authors) show the same 
trend, although with a high dispersion since it is a very small sample, and 
the tendency remains the same when 
adding them to our own sample, leading to the following linear relations: 

OG = (4.7$\pm$15.3)$\times$log10(d3D) + (4.00$\pm$9.0) for the early types

OG = (32.9$\pm$20.3)$\times$log10(d3D) + (-6.98$\pm$13.1) for the late types

\section{Discussion and conclusion}

We have shown from a selected homogeneous sample of 29 spiral galaxies that the
OG increases with 
the 3D cluster centric distance with a significant slope for late type
galaxies, meanwhile early ones show no peculiar tendency. 

Looking at the morphological types, one can see that the early types are
dominant in the inner part of the clusters, meanwhile late types are found 
in the outer parts (see open and filled circles on Fig.~1).

This segregation phenomenon between early type and late type spirals has 
been demonstrated by Adami et al. (1998) from the analysis of a sample of 
2000 galaxies in 40 clusters. It appears as the natural extension of the
segregation already well known for elliptical, S0 and spiral galaxies
in clusters (e.g. Melnick and Sargent, 1977; Dressler, 1980; Whitmore and 
Gilmore, 1991; Stein, 1997). 
This distribution of galaxies, depending on their type, may explain 
the observed tendency for OG to increase with the distance to the center of
a cluster, since there is a correlation between the morphological type 
and the shape of the RC, as shown by Corradi and Cappacioli (1990). 
They have shown indeed, from a sample of 167 galaxies, 
that flat curves are associated to the earlier types and rising curves to 
the late types. We confirm this effect when analysing the larger sample 
(245 galaxies) catalogued by Corradi and Capaccioli (1991)
for which they give $\simeq$ 200 shapes of RC. 

Fig.~2 shows the histogram of the distribution of types for the 73 rising RC
of that sample (there are only 14 decreasing curves, 
evenly distributed among the different types, and 112 flat curves for which
the histogram is practically the complement of Fig.~2). Among the 73 galaxies
with rising RC, 15 are found isolated and may be considered as field galaxies,
meanwhile 58 are found in groups or clusters.  
On Fig.~2 we plotted indeed two histograms side by side, one for the 51 field 
galaxies (hatched areas) and another one for the 148 cluster galaxies 
(gray areas).
The same trend is observed in both cases, namely that the percentage of 
rising curves clearly increases regularly from early to
late type spirals (we also checked that on our own sample of 39 cluster 
galaxies).

\begin{figure} 
\vbox 
{\psfig{file=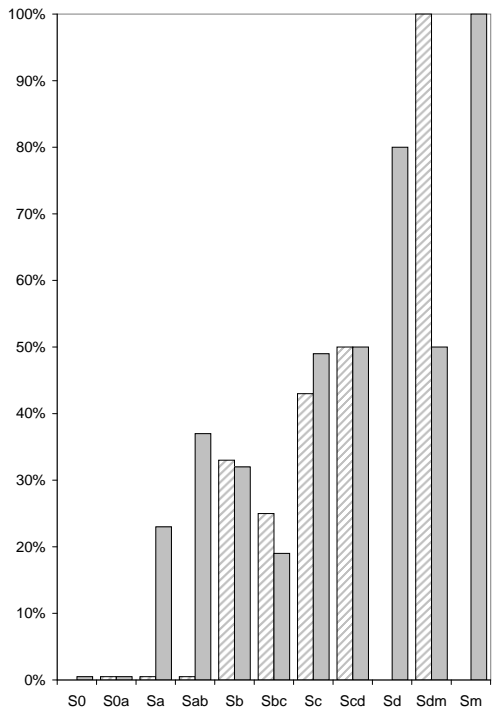,width=8.0cm,angle=0.}} 
\caption[]{Histogram of the distribution of galaxy types for the 73 rising 
rotation curves from the sample of Corradi and Cappacioli (1991). The percentage
of rising RC is plotted for each type. Two cases are distinguished : field
galaxies (hatched areas) and cluster galaxies (gray areas)}
\label{} 
\end{figure}

This suggests that the value of OG is more dependent 
on the type of a galaxy than on its evolution within a cluster.
We conclude that the correlation between OG and the morphological type of 
galaxies seen on Fig.~1 mainly reflects the importance of the dark halo, 
increasing with the type. 

Then the tendency of OG to increase with the distance to a cluster
center, although remaining a controversial subject, could mainly reflect the
morphological segregation of galaxies, those with larger values of OG
being found preferentially in the outer parts of the clusters. 

We now conjecture why there is a marked tendency of OG to
increase with the distance for late type spirals, meanwhile the
OG of early type spirals remains around zero.
 
Our results suggest that the effects of evolution within clusters predominantly 
affect late type galaxies, since the shape of their RC is more 
clearly dependent on their position within the cluster than for the early type
galaxies. 

We propose the following scenario:

Galaxies in a cluster are undergo to interactions which make them evolve. Those 
getting closer to the center will experience a larger number of interactions, 
thus losing a significant fraction of their halo and exhibiting less rising RC.
This is why OG is found to be smaller in the central part of the clusters. 
Finally, 
with time elapsing, galaxies gradually reach more stable orbits closer to the
cluster center meanwhile they evolve toward earlier types, as suggested by
the trend observed in both the distribution and the velocity dispersion 
profiles for the different morphological types by Adami et al. (1998). 
The early type galaxies have almost finished their evolution within the cluster,
now remaining on more circular orbits (also closer to the center) than the late 
type ones. Their behaviour is more homogeneous and they show no special trend.
The late type galaxies are still evolving, with those closer to the cluster 
center being more evolved and displaying smaller OG values, because they have
experienced a larger number of interactions. As time runs they will eventually
turn into early type galaxies.
 
A question remains however :
Early types being closer to the center should experience more interactions,
hence loosing more material from their halo and exhibiting more decreasing RCs.
On the contrary we have seen that their OG remains around zero and does not
show any significant variation with the distance to the cluster center.
This could be because they are moving in the denser parts of the cluster,
hence accreting material more or less compensating the fraction of halo lost 
during interactions. Another possible explanation is that we have no type 1 (Sa)
nor type 0 (S0a) galaxies in our sample (because the data we used are based on 
emission lines measurements), maybe preventing to see any tendency 
that could be more clearly seen with very early types.   
    
Besides explaining the tendency of OG to increase with 
the distance to the cluster center, the proposed scenario would also explain 
why the behaviour is different for early and late type galaxies.

Our severe selection criteria, looking for high quality RCs of cluster galaxies,
led us to work on a rather limited sample. The main result of our study, namely
that early type galaxies exhibit flat RCs whatever their location in the 
cluster, while late type ones have all the more rising RCs as they are found
further from the cluster center, needs however to be confirmed with larger
homogeneous samples.

\begin{acknowledgements}

{AC acknowledges the staff of the Dearborn Observatory for their hospitality 
during his postdoctoral fellowship. DR acknowledges the staff of the Anglo 
Australian Observatory for their hospitality during her postdoctoral 
fellowship, she thanks the french ministery of foreign affairs for its help
with a Lavoisier grant. The authors thank the referee for very useful and 
constructive comments. The authors thank M. Ulmer for a detailed reading of 
the revised manuscript.}

\end{acknowledgements}

\appendix

\section{Appendix: Discussion on the six interacting galaxies removed from
the sample}

Z 160-106: the interaction with its companion may explain the strange shape of 
the RC, with a very high value of OG, found in AmI ; also it is quite 
surprising to find H$\alpha$ emission in this type of object, which is the only
type -2 galaxy in our sample, making it all the more suspect. 

DC 47: probably an interacting pair as explained in AmIV.

NGC 6045: already at the limit of being excluded for its high inclination, 75 
degrees, it appears warped with a companion at the end of the eastern arm, see
AmI. 

NGC 6050 / IC 1179: they appear to be an interacting pair although they have
flat RCs. Since there is a 1500 km.s$^{-1}$ difference in systemic velocities, 
we suggest that it is maybe a superposition, see AmIII.

NGC 3861: has a close companion superimposed on a spiral arm, see AmIII.

\vfill 

\begin{thebibliography}{} 


\bibitem{} Adami C., Biviano A., Mazure A., 1998, A$\&$A 331, 439

\bibitem{} Amram P., le Coarer E., Marcelin M., et al., 1992, A$\&$AS 94, 
175: AmI 

\bibitem{} Amram P., Sullivan W., Balkowski C., et al., 1993, ApJ 403, 
L59: AmII

\bibitem{} Amram P., Marcelin M., Balkowski C., et al., 1994, A$\&$AS 103, 
5: AmIII

\bibitem{} Amram P., Boulesteix J., Marcelin M., et al., 1995, A$\&$AS 113, 
35: AmIV

\bibitem{} Amram P., Balkowski C., Boulesteix J., et al., 1996, A$\&$A 310, 
737: AmV

\bibitem{} Calberg R., Yee H., Ellingson E. et al., 1996, ApJ 462, 32

\bibitem{} Corradi R. and Capaccioli M., 1990, A$\&$A 237, 36

\bibitem{} Corradi R. and Capaccioli M., 1991, A$\&$AS 90, 121

\bibitem{} Courteau S., 1997, AJ 114, 2402

\bibitem{} Dale D.A., Giovanelli R., Haynes M.P. et al., 1997,
AJ 114, 455 

\bibitem{} Dale D.A., Giovanelli R., Haynes M.P. et al., 1998,
AJ 115, 418 

\bibitem{} Distefano A., Rampazzo R., Chincarini G., de Souza R., 1990,
A$\&$AS 86, 7

\bibitem{} Dressler A., 1980, ApJ 236, 351

\bibitem{} Isobe T., Feigelson E.D., Akritas M.G., Babu G.J., 1990, ApJ 364, 104

\bibitem{} Mathewson D.S., Ford L., 1996, ApJSS 107, 97

\bibitem{} Mathewson D.S., Ford L. and Buchhorn, 1992, ApJ 389, 5

\bibitem{} Melnick J., Sargent W., 1977, ApJ 215, 401

\bibitem{} Persic M., Salucci P., 1995, ApJSS 99, 501

\bibitem{} Press W., Teukolsky S., Vetterling W., Flannery B., 1992 "Numerical
Recipes" Cambridge University Press

\bibitem{} Rubin V., Whitmore B., Ford W., 1988, ApJ 333, 522: RWF

\bibitem{} Sperandio M., Chincarini G., Rampazzo R., de Souza R., 1995, A$\&$AS
 110, 279: Sp

\bibitem{} Stein P. 1997, A$\&$A 317, 670

\bibitem{} de Vaucouleurs G., de Vaucouleurs A., et al., 1989, the Third
Reference Catalog: RC3

\bibitem{} Whitmore B., Forbes, D., Rubin V., 1988, ApJ 333, 542: RWF 

\bibitem{} Whitmore B., Gilmore ., 1991, ApJ 367, 64

\end{thebibliography}
\end{document}